\pdfoutput=1
\documentclass[10pt,onecolumn]{article}
\usepackage{color}
\usepackage{graphicx}
\usepackage{amsmath,amssymb}
\usepackage{url}
\usepackage{physics}
\newcommand{\bb}{\mathbf}

\begin{document}

\title{Faraday shield dissipation in the drivers of SPIDER based on electromagnetic 3D calculations}
\author{D. L\'opez-Bruna$^1$, S. Denizeau$^2$, I. Predebon$^{2,3}$, \\ A. La Rosa$^2$, C. Poggi$^2$, P. Agostinetti$^{2,3}$}
\date{\footnotesize 
\begin{flushleft}$^1$ Laboratorio Nacional de Fusi\'on - CIEMAT, Madrid, Spain\\
$^2$ Consorzio RFX (CNR, ENEA, INFN, Universit\`a di Padova, Acciaierie Venete SpA), Padova, Italy\\
$^3$ Istituto per la Scienza e Tecnologia dei Plasmi, CNR, Padova, Italy\\ %, Corso Stati Uniti 4, 35127, Padova, Italy\\
\end{flushleft}}

\maketitle

\abstract
{\small 
SPIDER (Source for the Production of Ions of Deuterium Extracted from Rf plasma) is a full-scale prototype of the ITER NBI source. It is based on the concept of inductive coupling between radio-frequency current drive and plasma. Present three-dimensional (3D) electromagnetic calculations of stationary RF fields in SPIDER permit an evaluation of the power dissipation in its main constituents.  Taking experimental plasma parameters as input, we concentrate on the power dissipation in the copper-made Faraday shield lateral wall (FSLW) of the source for discharges with and without a static magnetic filter field. In agreement with our previous results and a first comparison with calorimetry data from the FSLW cooling circuit, the FSLW cylinder alone absorbs around 50\% of the available power for the studied plasma parameters. A hypothesized improvement of transport confinement may increase significantly the efficiency.}

\section{\label{sec:Introduction}Introduction}

The ITER Neutral Beam Test Facility (NBTF) \cite{Toigo2017The-PRIMA-Test-} is dedicated to test and optimize the Neutral Beam Injection (NBI) system of the ITER device with full-scale prototypes \cite{iter-physics-expert-group-on-EP-1999chapter-6,iter-physics-expert-group-on-EP2000chapter6}. Two main experiments are developed in the NBTF:  SPIDER is devoted to study the plasma source and beam extraction, while MITICA is a complete NBI device that should incorporate all improvements in its source after SPIDER experience. A third experiment, complementary to SPIDER, allows technological developments and optimized engineering solutions in support of the upcoming experimental campaigns \cite{mario2024optimizing-the-}.

\begin{figure}[htbp] %  figure placement: here, top, bottom, or page
   \centering
   \includegraphics[width=0.5\textwidth]{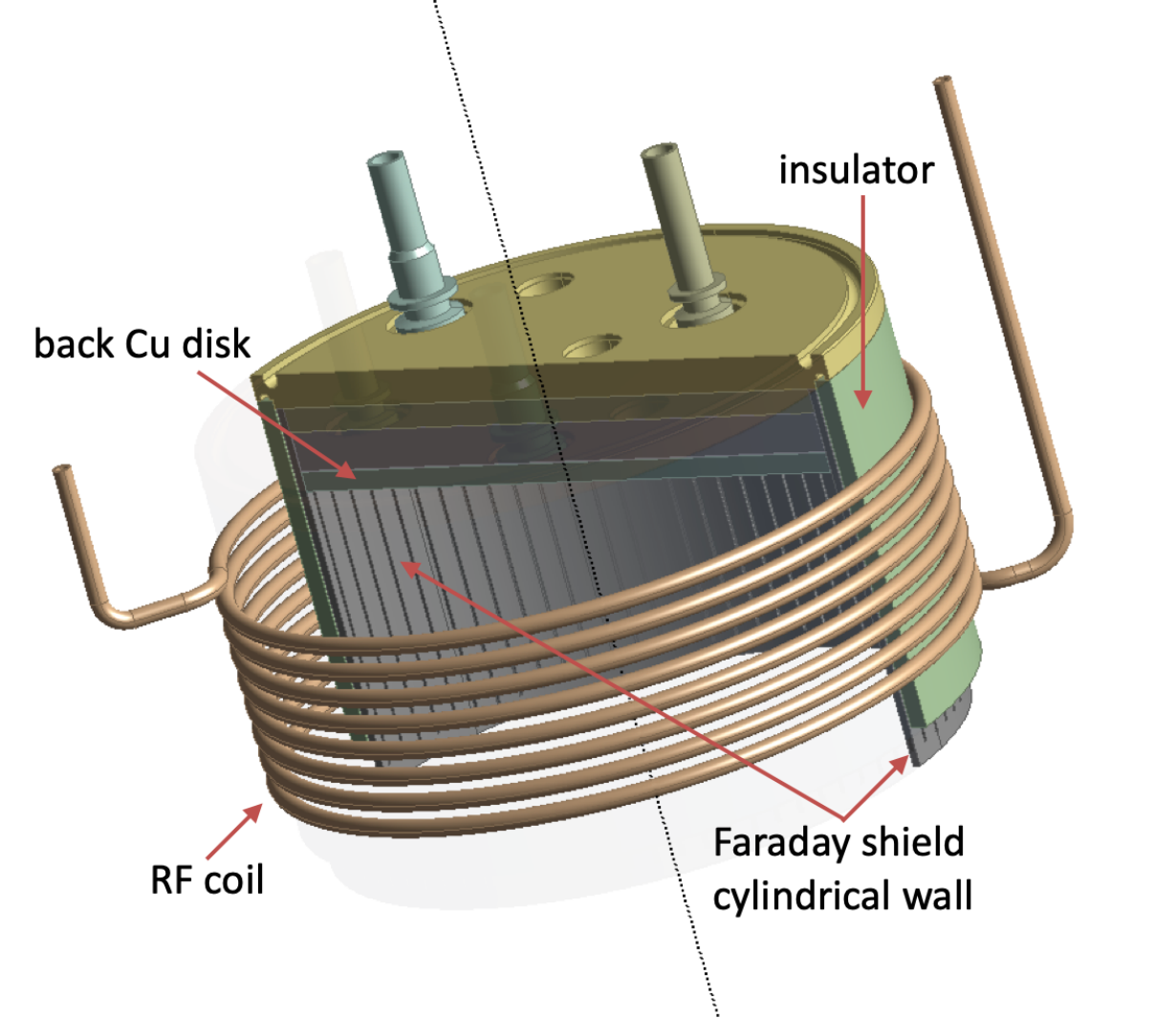} 
   \includegraphics[width=0.4\textwidth]{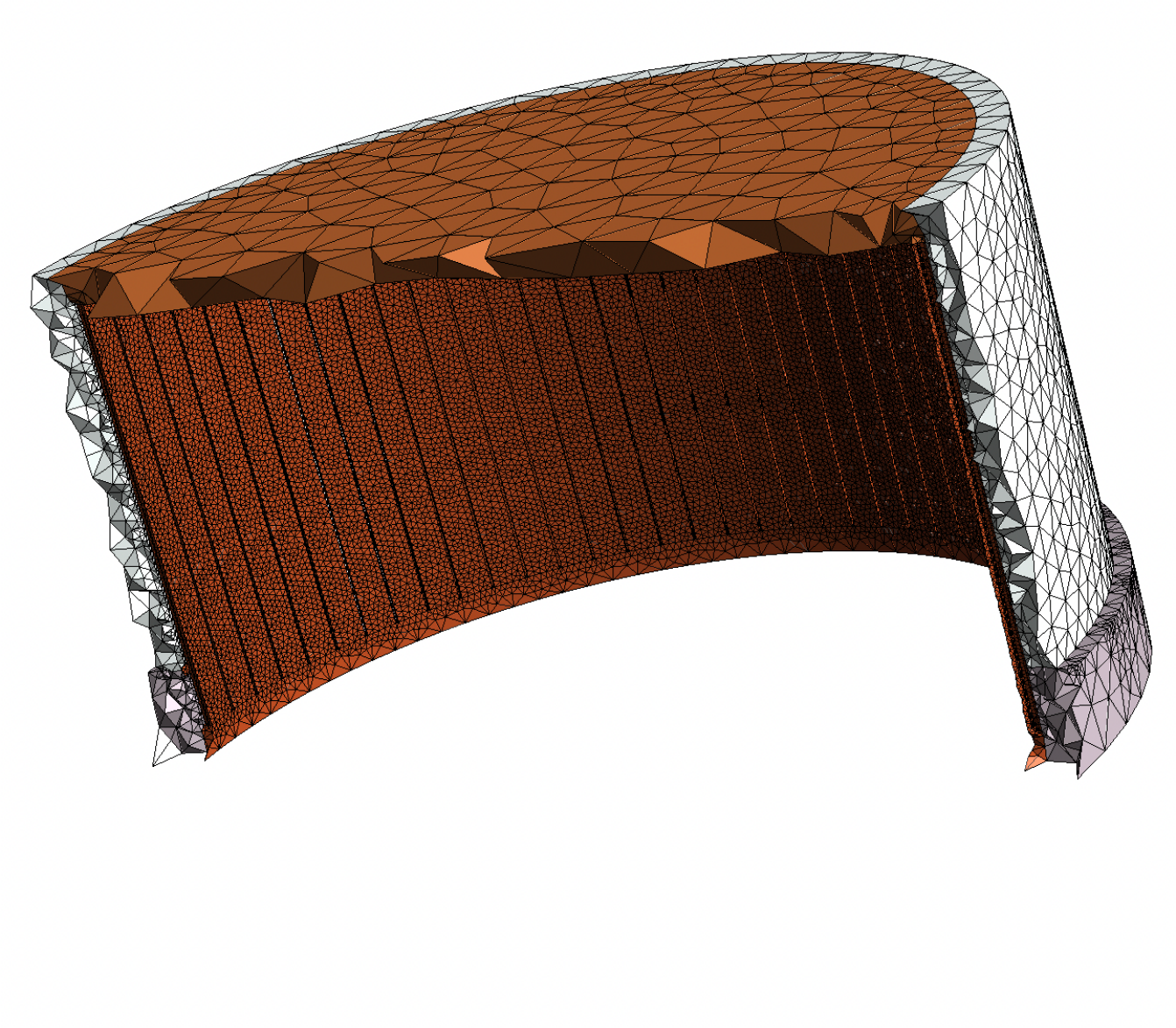} 
   \caption{CAD drawing of a SPIDER driver (left) and same cut view of the calculation mesh of the insulator and metallic parts (right). The bottom opens towards the plasma expansion region through a stainless-steel wall ---only a ring, shown in grey, belongs to the calculation domain.}
   \label{fig:real driver}
\end{figure}

The plasma source in SPIDER consists of a matrix of $2\times 4$ cylindrical cavities, commonly called ``drivers'' (see figure \ref{fig:real driver}, left), where $\approx 1$ MHz radio-frequency (RF) currents in the corresponding 8 RF-coils are coupled to the plasma via electro-magnetic induction. The plasma volume inside each driver is limited by a Mo-coated copper disk on the back and the cylindrical side of the Faraday shield lateral wall (FSLW), a water-cooled copper cylinder with longitudinal slits to allow for the penetration of the RF magnetic field while avoiding strong induced currents in the copper. The remaining circular side opens towards a larger chamber, common to the eight drivers, where the plasma expands until reaching the acceleration grids (see a complete description in \cite{toigo2021on-the-road-to-} and references therein).  An insulating case covers each driver and separates the FSLW from the RF-coil. 

SPIDER drivers are fed in groups of two by an RF generator with corresponding matching circuits. The electrical circuit associated to each pair of drivers can be described by an equivalent impedance, where the resistive part is related with the dissipated power at a given \emph{rms} current, $P=R_\mathrm{eff}I_{rms}^2$. This has allowed for experimental estimates of the effective resistances in different operating conditions \cite{jain2022investigation-o}. Likewise, different constituents of each driver can have an associated effective resistance such that $R_\mathrm{eff}$ is the sum of several effective resistances; for instance, those corresponding to the RF-coil or the FSLW. For a fixed RF-frequency, $R_\mathrm{eff}$ in presence of only linear conductors (e.~g.~the metallic parts) must not depend on the current $I_{rms}$ itself. The plasma, however, is expected to behave differently because, due to complicated transport processes, is not a linear conductor. Still, at a given RF-coil current, an effective resistance can be associated to the plasma as well, and therefore to the entire driver. As found in \cite{jain2022investigation-o}, different transport conditions give rise to different effective driver resistances. In particular, a strong experimental knob to alter plasma transport is the establishment of static magnetic fields \cite{toigo2021on-the-road-to-}. In SPIDER, their magnitude is controlled with the current $I_\mathrm{PG}$ through the plasma grid, complemented with a set of bus-bar conductors designed to obtain a desired filtering effect of the plasma electron flow towards the acceleration grids while minimizing the effect inside the drivers \cite{marconato2021an-optimized-an}. This has provided indeed a relevant result for the MITICA facility \cite{sartori2023highlights-of-r}. Static magnetic fields can also come from permanent magnets, which provide another element for possible optimization of the driver performance \cite{marconato2023integration-of-}.

Before being finally implemented in MITICA or, ultimately, in the ITER NBI systems, experimental results leading to changes in the design of the plasma sources should benefit from assessment from theory. One kind of theoretical study is the calculation of ohmic losses in the drivers, which include those in a Faraday shield that has been found to absorb a considerable fraction of the power delivered by the radio-frequency (RF) generator in these large sources \cite{briefi2022diagnostics-of-,lopezbruna2023threedimensional}. Ideally, a comprehensive transport model should be developed in order to understand the main physical reasons behind the behaviour of the plasmas in the source. Bidimensional calculations of transport including the electromagnetic coupling have been already developed for SPIDER \cite{zagorski20222-d-fluid-model,zagorski20232d-simulations}. On the other hand, an electromagnetic study of the sources that includes dissipation in the Faraday shield demands three-dimensional calculations due to its intrinsically 3D structure. Corresponding codes have been also developed for SPIDER. They have been checked numerically and used to estimate the effective resistances of the several parts of the driver, including model plasmas taken from experimental data; and also to study the importance of the geometry of the FSLW slits in the driver efficiency \cite{lopezbruna2023threedimensional,lopezbruna2022Three-dimensional}. In line with other studies \cite{chen2018electromagnetic,wang2022design-optimiza}, it seems that working with the design of the slits in the FSLW is not a promising way to increase the driver efficiency. On the other hand, numerical studies must be supported from the experiment, which is a very difficult task in large sources like SPIDER.

 In the present work we document our methods to obtain numerically the effective resistance of different metallic parts of the drivers, which are fundamental for the plasma transfer efficiency (PTE), or ratio between the plasma absorbed power and the power delivered by the RF generator. Additionally, we show the spatial distribution of the ohmic power density in the Faraday shield and make a first comparison with calorimetry data to confirm the high percentage of losses in the FSLW cylinder of the SPIDER drivers, roughly around half the generator power.  To avoid making too long a paper, we have separated the description of the thermal calculations and the calorimetry system in a companion work  \cite{denizeau2024heating}.  
 
  The rest of this paper is organized as follows: In Section \ref{sec:modelo} we describe briefly the calculations and the relevant model input/output. In Section \ref{sec:resultados} we explain the method we use to obtain the input RF-coil current for a given experimental plasma and RF-generator power (\ref{subsec:barridos}), discuss the driver efficiency and show the comparison with ohmic dissipation in the FSLW (\ref{subsec:FSLWdissipation}). The work finishes with the conclusions in Section \ref{sec:discusion}.

\section{\label{sec:modelo}Electromagnetic 3D model}

Figure \ref{fig:real driver} (left) shows the main elements of the driver relevant for the 3D electro-magnetic calculations: the insulator made in alumina, the copper disk and the cylindrical part of the FSLW. A cylindrical electromagnetic shield (not shown) covers the RF coil. Figure \ref{fig:real driver} (right) shows in coppery colour the calculation mesh of the back disk and the FSLW cylinder. The insulator is shown in white to ease the comparison with the left figure, but other parts of the calculation domain, like the plasma region, are not shown for clarity (but see \cite{lopezbruna2023threedimensional,lopezbruna2022Three-dimensional} for details).

\subsection{\label{subsec:ecuacion}Equation}

We solve the equation for the spatial dependence of the magnetic vector potential in the 3D geometry of the drivers of SPIDER considering (i) a single harmonic current in the driver RF coil, (ii) a practical formulation for the plasma conductivity \cite{Jain2018Evaluation-of-p,Jain2018Improved-Method,zagorski20232d-simulations}, and (iii) a calculation domain that excludes the RF winding current density, $\mathbf{J}_\mathrm{b}$. Point (iii) permits solving a homogeneous equation, $\mathbf{J}_\mathrm{b}=0$, but requires finding the appropriate boundary conditions dependent on the RF-coil current amplitude. They can be obtained more easily from 2D calculations assuming cylindrical symmetry near the RF coil (details about the calculations can be found in \cite{lopezbruna2022Three-dimensional} and its extended version \cite{lopezbruna2023threedimensional}). Thus, we solve the equation
\begin{equation}
\curl \curl \mathbf{A} (\mathbf{x}) + (\imath  \omega \mu \sigma - \omega^2 \mu \epsilon ) \mathbf{A} (\mathbf{x}) = \mu \mathbf{J}_\mathrm{b} = 0.
\label{ec:olla2_1}
\end{equation}
The notation is standard: $\mu$ and $\epsilon$ are, respectively, the magnetic permeability and dielectric constant of the materials. Since we do not deal with magnetic materials, we take the vacuum magnetic permeability. In the case of $\epsilon$, we adopt the vacuum value except for the alumina insulator (figure \ref{fig:real driver}). With this exception, the electromagnetic properties of the materials are described via their conductivity $\sigma$.

Equation \ref{ec:olla2_1} is solved using the Finite Element Method (FEM) with the numerical tools offered by the FEniCS project \cite{fenicsproject,Logg2012Automated-Solut,Langtangen2017Solving-PDEs-in}, which provides an appropriate language to manage integral forms \cite{Alnaes2012UFL} and a large set of tools to code FEM problems \cite{Logg2010DOLFIN,Logg2012DOLFIN}. The 3D mesh, visible in figure \ref{fig:real driver} for the Faraday shield and insulator, has been built with the Gmsh software \cite{Geuzaine2009Gmsh:-A-3-D-fin}.

The driver geometrical model is described in \cite{lopezbruna2023threedimensional,lopezbruna2022Three-dimensional}. Here we recall that the mesh is refined near the FSLW to account for the large conductivity of copper, $\sigma_\mathrm{Cu} \gtrsim 4\cross 10^{7}$ S/m at the temperatures of interest, and the corresponding tiny skin depth, $\sim 0.1$ mm at the relevant RF. This has required $\sim 10^6$ cells and order-two vector function families. 

The calculations can be reduced to a smaller domain, like the plasma region alone. This has been used to check their validity, as done in \cite{lopezbruna20212Delectromagnet}, when there is a known solution to compare with (e.~g.~in vacuum taking the boundary conditions of an infinite solenoid or a set of discrete circular coils.) In the presence of plasma, 3D calculations have also been satisfactorily compared with equivalent 2D calculations that use very different numerical methods \cite{lopezbruna2023threedimensional,lopezbruna2022Three-dimensional}. More complex calculations involving metallic parts are exposed in what follows.

\subsection{\label{subsec:inputs}Model inputs}

The calculations solve equation \ref{ec:olla2_1} considering the electromagnetic properties of the materials in the calculation domain (insulator, copper, plasma, vacuum), the angular frequency of the RF-drive, the boundary conditions and the possible presence of a static magnetic field. In the present calculations the plasma electron density and electron temperature are also taken as external inputs from the experimental information. There are, however, two main inputs that are far from immediate to obtain: the plasma electrical conductivity and the amplitude of the current in the RF coil, $I_\mathrm{RF}$.

\begin{figure}[htbp] %  figure placement: here, top, bottom, or page
   \centering
   \includegraphics[width=0.93\columnwidth]{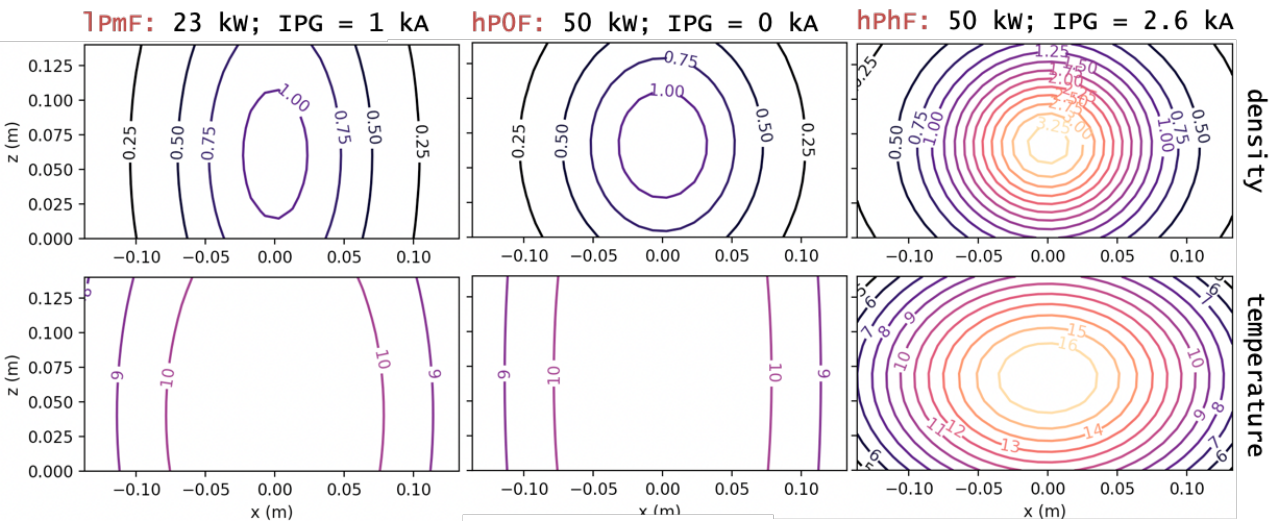}
   \caption{Contour plots of the input electron density (up) and temperature (down) adopted for SPIDER discharges operated at the indicated values of nominal RF power and plasma grid current. The labels, in red, stand for \emph{low power--moderate filter field} (lPmF), \emph{high power--null filter field} (hP0F) and \emph{high power--high filter field} (hPhF). See table \ref{T:plasmas}.}
   \label{fig:perfiles}
\end{figure}

The plasma electrical conductivity, $\sigma$, is a function of magnitudes like the electron density and temperature and the neutral gas pressure. Unless a transport problem that includes momentum balance for the charged plasma species is solved consistently with Eq.~\ref{ec:olla2_1}, we must use some suitable formulation for the plasma conductivity. Simplified versions of the plasma conductivity have turned out to yield acceptable values of the power absorbed by the plasma in previous works \cite{Jain2018Improved-Method,zagorski20232d-simulations,lopezbruna2023threedimensional,lopezbruna2022Three-dimensional,lopezbruna20212Delectromagnet,jain2022investigation-o,zagorski20222-d-fluid-model}. For this reason, despite their lack of detailed accounting of non-linear and non-local effects, the conductivity models used in all these previous works for SPIDER are appropriate for the present 3D calculations. Still, we need reliable distributions of the electron density and temperature inside the driver. During y.~2020, some SPIDER campaigns (e.~g.~campaign S16) were devoted to obtain plasma parameters with spectroscopy, movable electric probes \cite{sartori2021development-of-} and other diagnostics \cite{lopezbruna20212Delectromagnet}. In this work we use three sets of plasma data, which correspond to discharges operated under different nominal power, $P_\mathrm{RF}$, and plasma grid current, $I_\mathrm{PG}$. Figure \ref{fig:perfiles} shows contours of the distributions adopted for these plasmas. We observe that $I_\mathrm{PG}$ affects transport, making the profiles more peaked and dense. The shown contours are approximations based on the available experimental data. More information on these plasmas is given in Table \ref{T:plasmas}.

With respect to the electrical conductivity of metallic materials present in the calculation domain, it is enough providing a reasonable value of their working temperature. Uncertainties in this temperature convey uncertainties in the estimated dissipated powers, as discussed below.

\begin{table}[htp]
\caption{Relevant discharge characteristics of the SPIDER plasmas used in this work: nominal power per driver, $P_\mathrm{RF}$; current in the plasma grid (magnetic filter field), $I_\mathrm{PG}$; and hydrogen gas pressure, $p_\mathrm{g}$. The last two columns show the calculated current amplitude in the RF coils, $I_\mathrm{RF}$, and the labels used to identify each type of input plasma (figure \ref{fig:perfiles}.)}
\begin{center}
\begin{tabular}{ccccc}
$P_\mathrm{RF}$ & $I_\mathrm{PG}$ & $p_\mathrm{g}$ & calc.~$I_\mathrm{RF}$ & label\\
(kW) & (kA) & (Pa) & (A) &\\
\hline\hline
23 & 1 & $0.35$ & 180 & lPmF\\
50 & 0 & $0.34$ & 250 & hP0F \\
50 & $2.6$ & $0.34$ & 230 & hPhF \\
%30 & $1$ & $0.30$ & 210 \\
\end{tabular}
\end{center}
\label{T:plasmas}
\end{table}

$I_\mathrm{RF}$ is not an available measurement in a large plasma source like SPIDER. In essence, it must be inferred from other experimental data like the nominal net power delivered to the driver and the plasma characteristics. A good part of this paper is devoted indeed to identify consistent values for $I_\mathrm{RF}$ in the plasma conditions of figure \ref{fig:perfiles}, which we collect in table \ref{T:plasmas}. As will be explained below, we can use $I_\mathrm{RF}$ as a variable parameter, which, together with information on the plasma, will be fixed to the value that justifies the nominal power.  Table \ref{T:plasmas} gives the identified $I_\mathrm{RF}$ values.

\subsection{\label{subsec:salidas}Model output}

The main output of the calculations is the spatial distribution of the magnetic vector-potential field, $\vb{A}$. Some immediate derived fields are the induced magnetic, $\vb{B} = \curl \vb{A}$, and electric field, $\vb{E} = - \imath \omega \vb{A}$. From the latter we obtain the effective current density,
\begin{equation}
\vb{J}_\mathrm{eff} =  \frac{1}{\sqrt{2}}\Re \{-\imath \omega \sigma \vb{A}\} =  \frac{1}{\sqrt{2}} \omega (\sigma_r \bb{A}_i + \sigma_i \bb{A}_r)
\label{ec:olla36_3}
\end{equation}
and the root-mean-square value of the ohmic power,
\begin{equation}
P_\Omega^\mathrm{rms} =  \frac{1}{2} \Re \left \{ \int_\mathrm{ID} \vb{E} \cdot ( \sigma \vb{E})^* \dd x \right \} =  \frac{1}{2}  \int_\mathrm{ID}  \omega^2 \sigma_r (A_r^2 + A_i^2)  \dd x;
\label{ec:Prms}
\end{equation}
where we have indicated with sub-indexes the real and imaginary parts of the different complex magnitudes and the asterisk indicates complex-conjugate. The volume integrals extend to the relevant integration domain (ID), which could be the plasma or the different copper parts. Incidentally, this integral is always negligible in the insulator part shown in figure \ref{fig:real driver}.

\section{\label{sec:resultados}Results}

The vacuum inductance of SPIDER drivers has been estimated in previous works, $L \approx 9.5$ $\mu$H \cite{jain2022investigation-o}. It is known that $L$ changes very little in presence of the plasma, a fact that was the base to deduce that the electrical conductivity of SPIDER plasmas must be strongly reduced in presence of both, RF and static magnetic fields \cite{lopezbruna20212Delectromagnet}. Considering that the RF coil is a set of eight single perfectly circular coils, the present 3D electromagnetic calculations yield a vacuum driver inductance  $L_\mathrm{d}^\mathrm{vac} =  9.55$ $\mu$H with very little variation (always towards lower values) in presence of the plasma.

\subsection{\label{subsec:barridos}RF-coil current and effective resistances}

Equation \ref{ec:Prms} is used to evaluate the power dissipated in the calculation sub-domains, e.~g.~the different metallic parts or the plasma. As mentioned at the end of Sec.~\ref{subsec:inputs}, the RF-coil current is a calculation parameter. Thus, for a given plasma, we perform $I_\mathrm{RF}$ scans from which we obtain a set of pairs $(I_\mathrm{RF}, P)$, where $P$ represents the ohmic power absorbed in the different sub-domains. Then we estimate the corresponding effective resistances from fits to a power law,
 \begin{equation}
P = \alpha I_\mathrm{RF}^\beta,
\label{ec:powerlaw}
\end{equation}
where we expect $\beta \approx 2$ for the metallic parts, like the FSLW. Since $I_\mathrm{RF}$ is the amplitude of the RF current,  the effective resistance $R_\mathrm{eff} \equiv 2P/I_\mathrm{RF}^2$ must be taken as $R_\mathrm{eff} = 2\alpha$.

\begin{figure}[h] %  figure placement: here, top, bottom, or page
   \centering
   \includegraphics[width=0.73\columnwidth]{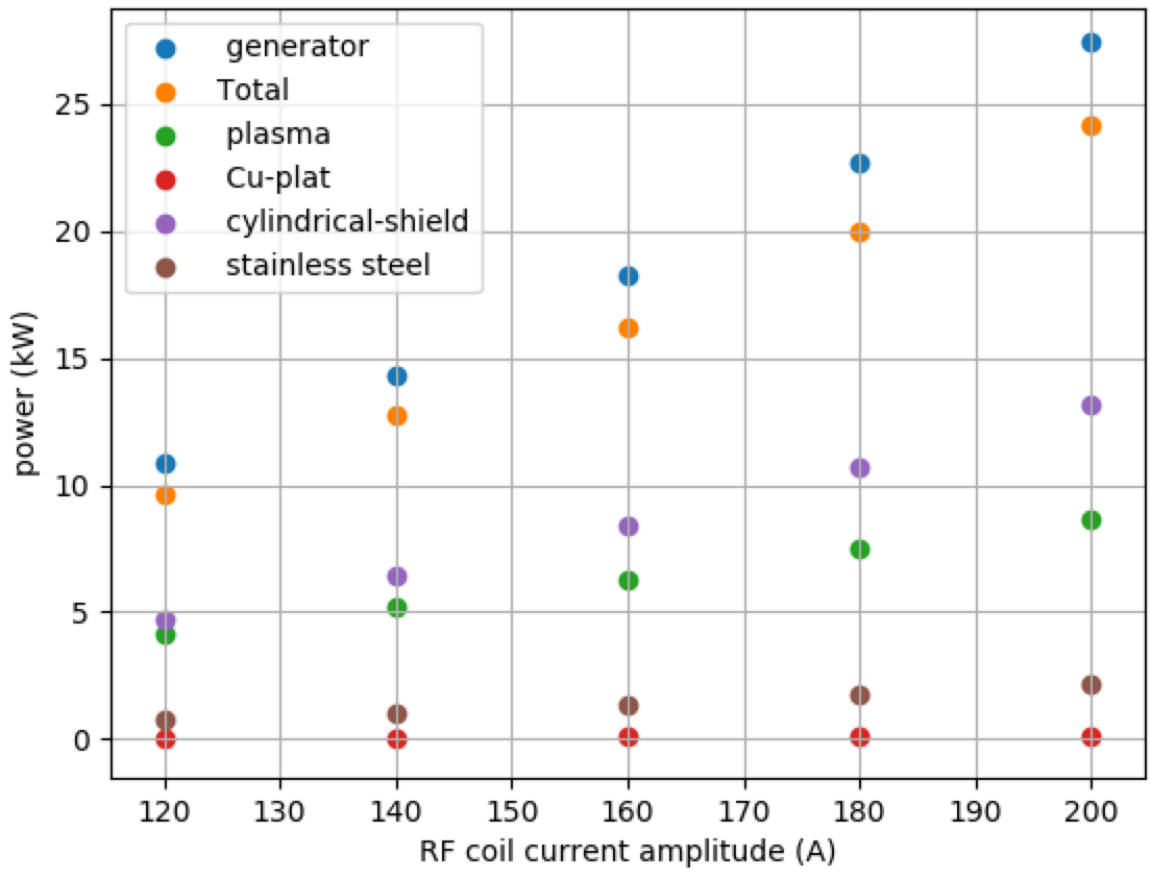}
   \caption{$I_\mathrm{RF}$ scan for lPmF plasmas (table \ref{T:plasmas}) using $\sigma_\mathrm{Cu}=5 \times 10^7$ S/m. The ``Total'' value is the sum of all calculated values (plasma + back disk + FSLW + stainless steel). The ``generator'' power adds also the losses in the RF coil and the electromagnetic shield.}
   \label{fig:Ibscan1}
\end{figure}

Figure \ref{fig:Ibscan1} is an example of $I_\mathrm{RF}$ scan. We underline that scans like this are done with \emph{fixed} plasma characteristics, which is not consistent because the real plasmas could depart considerably from the assumed input plasmas when $I_\mathrm{RF}$ is changed.  Note, however, that this is not a concern for the metallic parts because the power they dissipate depends very little on the characteristics of the plasma for a given $I_\mathrm{RF}$ \cite{lopezbruna2023threedimensional,lopezbruna2022Three-dimensional}.  Therefore, we can safely extend the scans to obtain effective resistances of the metallic parts, and later use the information of the nominal power to infer $I_\mathrm{RF}$ values for each of the three plasmas considered.

The scan presented in figure \ref{fig:Ibscan1} corresponds, in particular, to the lPmF input plasma with $P_\mathrm{RF} = 23$ kW, gas pressure $p_\mathrm{g} \approx 0.3$ Pa and filter magnetic field current $I_\mathrm{PG}= 1$ kA (table \ref{T:plasmas}). 
Based on equation \ref{ec:Prms}, the calculations yield the ohmic power dissipated in the stainless steel part of the domain, which encircles the aperture of the driver to the expansion zone of the source (brown); the power dissipated in the copper cylindrical part of the FSLW (purple), the copper disk (red) and the plasma (green).  The sum of all these terms are added to give a ``Total'' (orange). The generator power is then estimated by adding the power dissipated by two elements not included in the calculation domain: the RF-coil and the cylindrical electro-magnetic shield, a metallic structure that covers the RF winding. Since their effective resistances are calculated separately \cite{lopezbruna2023threedimensional,lopezbruna2022Three-dimensional,jain2022investigation-o}, the power dissipated in these two elements can be simply added as proportional to $I_\mathrm{RF}^2$. The consequent generator power is shown with blue dots in figure \ref{fig:Ibscan1}. The reader can check that the experimental value of the generator power, 23 kW in this case, is justified when $I_\mathrm{RF} \approx 180$ A. This is the value we have written in table \ref{T:plasmas}. 

\begin{table}[h]
\caption{Effective resistance of the stainless steel, copper parts (FSLW cylinder + disk) and two-drivers equivalent resistance, $R_\mathrm{D}$ with plasma, based on coil-current scans like the one in figure \ref{fig:Ibscan1}, for the indicated plasmas.}
\begin{center}
\begin{tabular}{lccc}
plasma type & $R_\mathrm{eff,ss}$ (m$\Omega$) & $R_\mathrm{eff,FS}$ (m$\Omega$)  & $R_\mathrm{D}$  ($\Omega$)\\
\hline\hline
lPmF & 103 & 648 & $2.8$\\
hP0F & 103 & 654 & $3.2$\\
hPhF & 99 & 636 & $3.8$\\
\end{tabular}
\end{center}
\label{T:LIBERTY103_2}
\end{table}

 Scans similar to the one of figure \ref{fig:Ibscan1} have been done with the other two types of plasma shown in figure \ref{fig:perfiles}. The exponent found for all the metallic parts is always $\beta = 2$ within a 1\%, which yields the estimates of effective ohmic resistances shown in  Table \ref{T:LIBERTY103_2}. The model Faraday shield, composed of the FSLW cylinder and the copper back disk, has an effective resistance  $R_\mathrm{eff,FS} \approx 0.65$ $\Omega$, quite larger than the values found for the stainless steel, $R_\mathrm{eff,ss} \approx 0.1$ $\Omega$ or the electromagnetic shield,  $R_\mathrm{eff,EM} \approx 0.044$ $\Omega$ \cite{lopezbruna2023threedimensional,lopezbruna2022Three-dimensional}. Adding the resistance of the RF-coil, $R_\mathrm{eff,b} \approx 0.12$ $\Omega$ \cite{jain2022investigation-o}, we estimate the effective resistance of the passive (metallic) elements in a single SPIDER driver during normal plasma operation, $R_\mathrm{eff} \approx 0.9$ $\Omega$. 
 
 Note that $R_\mathrm{eff}$ above is \emph{not} the driver equivalent resistance, which we can approach as $2P_\mathrm{RF}/I_\mathrm{RF}^2$ per driver using the values in table \ref{T:plasmas}. To ease the comparison with available measurements of the equivalent driver resistance (referred to each generator, i.~e.~two drivers in series) \cite{jain2022investigation-o} we have written the obtained values also in table \ref{T:LIBERTY103_2}. We must warn that, in  \cite{jain2022investigation-o}, only a limited set of discharges could be used and the operating conditions were not identical to those used for table \ref{T:LIBERTY103_2}. In addition, the measurements were based on the active power delivered by the generator (measured with better accuracy than previously, as it pertains to our calculations) and the \emph{rms} current at its output; whereas our estimates of $R_\mathrm{D}$ assume the obtained $I_\mathrm{RF}$ from the RF-coil and specific passive components, assumed dominant in terms of power dissipation (e.~g.~there is no transmission line). Still, we propose two cases for comparison: for a source pressure $p_\mathrm{g}=0.38$ Pa at an RF power of 45 kW/driver in hydrogen and $I_\mathrm{PG}=2.5$ kA (approximately similar to our hPhF case), the authors in \cite{jain2022investigation-o} found $R_\mathrm{D} \approx 4.2$ $\Omega$. In plasmas operated at $p_\mathrm{g}=0.40$ Pa at an RF power of 29 kW/driver in hydrogen and $I_\mathrm{PG}=1.0$ kA (not too far from our lPmF case), they found  $R_\mathrm{D} \approx 3.7$ $\Omega$. Both values are larger than our estimates in table \ref{T:LIBERTY103_2}, but in acceptable agreement given the uncertainties.

\subsection{\label{subsec:eficiencia}Driver efficiency}

Power-law coefficients for a given input plasma can be used to estimate the driver efficiency. 
 From equation \ref{ec:powerlaw}, we can associate coefficients to the plasma and generator power delivered to one driver; respectively $P_\mathrm{p} = \alpha_\mathrm{p} I_\mathrm{RF}^{\beta_\mathrm{p}}$ and $P_\mathrm{G} = \alpha_\mathrm{G} I_\mathrm{RF}^{\beta_\mathrm{G}}$. By definition, the efficiency is their ratio,
 \begin{equation}
\eta = \frac{\alpha_\mathrm{p}}{\alpha_\mathrm{G}} I_\mathrm{RF}^{\beta_\mathrm{p}-\beta_\mathrm{G}}.
\label{ec:eficiencia}
\end{equation}
The current that corresponds to a given power of interest is also obtained from the fit Eq.~\ref{ec:powerlaw}. Let $P_X$ be given, where $X$ may refer, for example, to the plasma ($X \equiv$ p) or the generator ($X \equiv$ G). Then
 \begin{equation}
I_\mathrm{RF} = \left ( \frac{P_X}{\alpha_X} \right )^{1/\beta_X}.
\label{ec:Ib dada Ppl}
\end{equation}
Equation \ref{ec:Ib dada Ppl} provides $I_\mathrm{RF}$ after the plasma data and $P_\mathrm{G}$ are taken from the experiments. Observe, however, that its predictive value is limited to plasmas not too different from the one used for the fit. Thus, equation \ref{ec:Ib dada Ppl} can give an indication of the $I_\mathrm{RF}$ necessary for a guessed plasma absorbed power $P_\mathrm{p}$, as long as the plasma can be considered similar to the experimental one used as input for the fit.

\begin{figure}[htbp] %  figure placement: here, top, bottom, or page
   \centering
   \includegraphics[width=0.6\columnwidth]{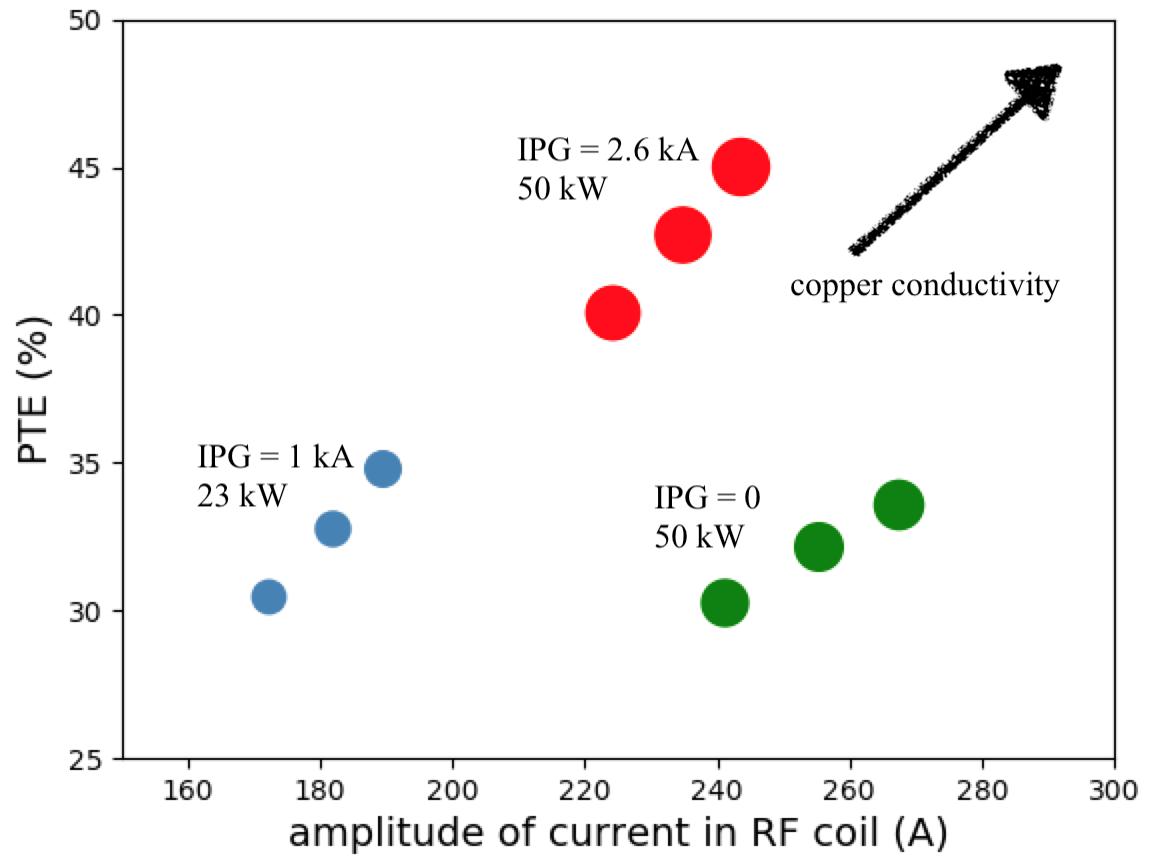} 
   \caption{Power transfer efficiency in percentage, PTE $\equiv 100\eta$, obtained for each type of plasma in table \ref{T:plasmas} and increasing values of copper conductivity.}
   \label{fig:eficiencias}
\end{figure}

 The dissipation in the metallic parts depends on their electrical conductivity, in turn dependent on the temperature. Next we evaluate the sensitivity of the dissipation in the FSLW cylinder to copper temperature. So far we have assumed $\sigma_\mathrm{Cu} = 5 \times 10^7$ S/m, which corresponds to around $50$ $^\circ$C. Now we use formulae \ref{ec:eficiencia} and \ref{ec:Ib dada Ppl} to produce figure \ref{fig:eficiencias}, where $I_\mathrm{RF}$ scans have been repeated for each type of plasma in table \ref{T:plasmas} using three copper conductivity values, $\sigma_\mathrm{Cu} \in \{4, 5, 6\} \times 10^7$ S/m. These  correspond, roughly, to respective temperatures $\{100, 50, 10\}$ $^\circ$C. The hotter the copper, the lower the $\sigma_\mathrm{Cu}$ and the larger the power dissipated in the Faraday shield for a given generator power; in other words, lower efficiencies correspond to lower $\sigma_\mathrm{Cu}$ and conversely. With higher electrical conductivity, also higher $I_\mathrm{RF}$ is required to justify a given RF power because the scaling of the power deposited in the plasma is weaker than the quadratic law for the metallic parts (see figure \ref{fig:Ibscan1}). Observe that the relevant copper temperatures are those in the regions where electric conduction concentrates, which are expected to be the hottest parts of the Faraday shield depending on the balance between electric induction and active water cooling. This aspect is further discussed in the companion paper \cite{denizeau2024heating}.

The efficiencies shown in figure \ref{fig:eficiencias} are to be considered approximate. The plasma conductivity models are taken from simplified practical formulations \cite{Vahedi1995Analytic-model-,tuszewski1997inductive-elect}. 
 At the same time, there are uncertainties in the measurements of plasma properties (including the geometrical characterization of profiles). Therefore, we interpret figure \ref{fig:eficiencias} in semi-quantitative terms: the PTE in the drivers of the SPIDER prototype is around 30--45\% depending on the presence of a filter field, which increases the efficiency; the RF-coil currents, assuming Cu temperatures near 50 $^\circ$C in the conduction zones, should be around the values written in table \ref{T:plasmas}; and a cooler (or more electrically conductive) Faraday shield increases the driver efficiency.

\begin{table}[htbp]
\begin{center}
\caption{PTE based on the regular plasma shapes of figure \ref{fig:perfiles} (right) or with doubled density and reduced electron temperature with normal ($18.4$ cm) or reduced ($17.4$ cm) radius of the electro-magnetic shield. The values are obtained at fixed generator power, $P_\mathrm{RF}$ and fixed plasma absorbed power, $P_\mathrm{pl}$. In the last case, the coil-current amplitude is also shown.}
\label{T:LIBERTY95_1}
\begin{tabular}{lcccc}
profiles & radius (cm) & $P_\mathrm{RF} = 47$ kW & $P_\mathrm{pl} = 20$ kW: & $I_\mathrm{RF}$ (A)\\
\hline\hline
regular & $18.4$ &  43\% & 43\% & 258\\
regular & $17.4$ &  40\% & 39\% & 305\\
doubled & $17.4$ & 53\% & 56\% & 227\\
\end{tabular}
\end{center}
\end{table}

The present numerical tools permit also inferring that the efficiency changes considerably if the plasma transport properties are modified, for instance via some appropriate configuration of static magnetic fields. Indeed, new experiments in SPIDER and its satellite experiment MINION \cite{mario2024optimizing-the-} foresee the test of new driver configurations with smaller-radius electromagnetic shield to allow for the insertion of permanent magnets around the cylindrical side of the FSLW. The electro-magnetic shield would have its radius reduced by 1 cm, which reduces the vacuum magnetic flux through the driver section to an 80\% of the original design. The added magnets, on the other hand, are expected to favour denser plasmas.
 
 A different electromagnetic shield changes the boundary conditions for our 3D calculations. To have an idea about how the reduced shield could impact the PTE, we impose (i) the hPhF profiles of figure \ref{fig:perfiles} and (ii) the same profile shapes with \emph{doubled} density profile and lowered peak electron temperature from 17 eV to 14 eV. Table \ref{T:LIBERTY95_1} shows that, for a plasma power deposition of 20 kW, the PTE increases from (i) 39\% to (ii) 56\%. The increased PTE under the assumption of improved confinement compensates the reduction of vacuum magnetic flux, since the original electro-magnetic shield with regular profiles gives a 43\% PTE for this plasma power absorption. The same qualitative results are obtained at fixed generator power. Evidently, a proper treatment of this study demands solving a transport problem. We leave these results as indication that the efficiency of the drivers depends essentially on the plasma transport properties.

\begin{table}[ht]
\centering
\caption{Relative losses in the FSLW cylinder when its electrical conductivity and thickness are changed at fixed  $I_\mathrm{RF}=200$ A. Reference cases have $\sigma_{50} = 5 \times 10^7$ S/m and $\delta_r^\mathrm{F} = 0.33$ cm. Shown are: the effective values of FSLW dissipated power, $P_\mathrm{F}$, and generator power, $P_\mathrm{G}$; the percentage of losses in the FSLW, $100P_\mathrm{F}/P_\mathrm{G}$, and its effective resistance.}
\label{T:LIBERTY148_1}
\begin{tabular}{cccccc}
  conductivity & thickness & $P_\mathrm{F}$ &  $P_\mathrm{G}$  & losses & $R_\mathrm{ef}$\\
  &  & (kW) & (kW) & (\%) & ($\Omega$) \\
\hline\hline
 $\sigma_{50}$  & $\delta_\mathrm{F}$ & $13.19$  & $22.01$ & $60$ & $0.66$ \\
 $0.77\sigma_{50}$  & $\delta_\mathrm{F}$ & $16.53$ & $25.37$  & $65$ & $0.83$ \\
$\sigma_{50}$  & $1.7\delta_\mathrm{F}$ & $13.67$  & $22.98$ & $59$ & $0.68$ \\
$0.77\sigma_{50}$  & $1.7\delta_\mathrm{F}$ & $17.33$ & $26.65$ & $65$ & $0.87$ \\
\end{tabular}
\end{table}

As a final example of efficiency estimates, we consider the possibility of having a thicker FSLW. At present there are advantageous technologies that may permit faster and more economic manufacturing processes, like additive manufacturing \cite{agostinetti2023innovative}. Ongoing research based on the SPIDER design is considering the possibility of using these techniques, but the final product would likely be thicker and possibly more resistive. Since the power absorbed by the FSLW at a given RF-coil current depends little on the plasma \cite{lopezbruna2023threedimensional,lopezbruna2022Three-dimensional}, we explore the variation of the FSLW conductivity and thickness in vacuum at fixed $I_\mathrm{RF} = 200$ A. Figure \ref{T:LIBERTY148_1} already tells us that the effect of the copper conductivity can easily make a difference of a 10\% in PTE. Let us assume the resistivity of the FSLW material augmented by a 30\%; or, equivalently, let us reduce the plasma conductivity by a factor $0.77$ taking as reference the conductivity of pure copper at a temperature near 50 $^\circ$C, $\sigma_{50}= 5\times 10^7$ S/m. This would be approximately the expected conductivity of CuCrZr, one of the materials proposed for additive manufacturing, at the same temperature. Likewise, let us consider the regular thickness of the FSLW cylindrical side, $\delta_\mathrm{F}=0.33$ cm, and a 70\% thicker wall, $1.70\delta_\mathrm{F}$. Table \ref{T:LIBERTY148_1} compiles the results, where it can be appreciated that the effect of the FSLW electrical resistivity is stronger than that of the augmented thickness. As before, a detailed study should be done with a proper evaluation of transport in the new conditions.

\subsection{\label{subsec:FSLWdissipation}Ohmic dissipation in the FSLW}

The low efficiencies found in section \ref{subsec:barridos} are due to the high effective resistance of the FSLW. As exemplified in figure \ref{fig:Ibscan1}, this metallic part may absorb about half the power delivered by the generator. Here we show how these heating losses are distributed in its cylindrical part.

The net power delivered by each of the four SPIDER RF-generators (half of which goes to one driver) is measured with acceptable accuracy \cite{jain2022investigation-o}. From the net power we can infer, using appropriate plasma profiles and Eq.~\ref{ec:Ib dada Ppl}, the coil current $I_\mathrm{RF}$ from which the power share can be obtained by the electro-magnetic calculations.  Unfortunately,  there were no dedicated measurements with movable probes during calorimetry experiments and only some central values of electron density and temperature are available. According to the data from the S20 campaign (y.~2021), however, there are plasmas with amenable calorimetry data and operational parameters $I_\mathrm{PG} = 1$ kA, $p_\mathrm{gas}=0.30$ Pa and $P_\mathrm{G} = 30$ kW, not very different from the lPmF type in table \ref{T:plasmas}, while the available central values of density and temperature can be considered equal within uncertainties. Consequently, we have adopted the plasma shapes from the S16 campaign (figure \ref{fig:perfiles}, left) and repeated the calculations imposing the current predicted by Eq.~\ref{ec:Ib dada Ppl} for a net $P_\mathrm{G} = 30$ kW, which results in $I_\mathrm{RF} = 210$ A.

\begin{figure}[htp] %  figure placement: here, top, bottom, or page
   \centering
   \includegraphics[width=0.69\columnwidth]{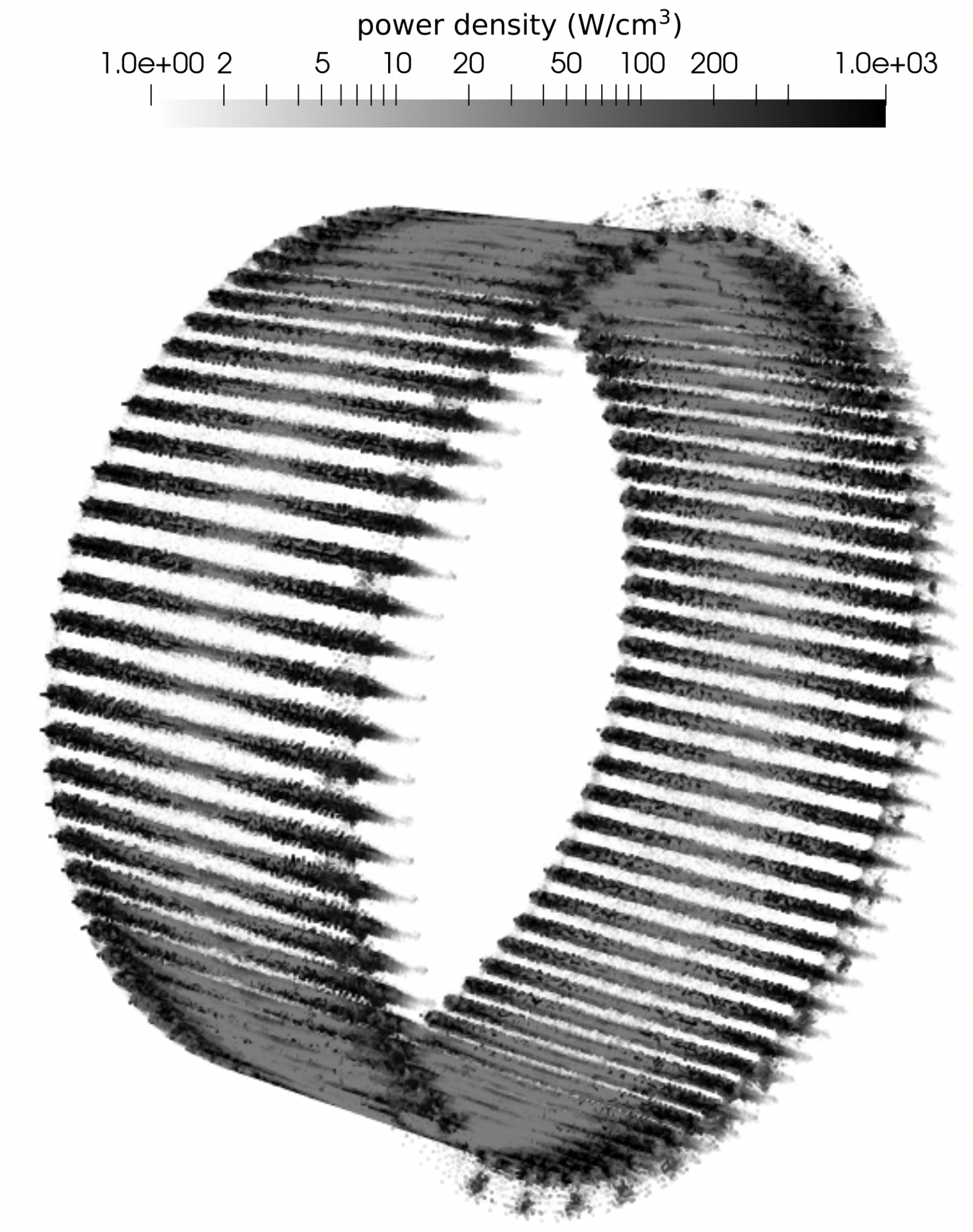} 
   \caption{Calculated ohmic power density distribution in the Faraday shield considering electron density and temperature distributions of a SPIDER plasma operated in hydrogen with $P_\mathrm{G} = 30$ kW per driver (estimated $I_\mathrm{RF}=210$ A), weak filter magnetic field ($I_\mathrm{PG} = 1$ kA) and low gas pressure, $p_\mathrm{gas}=0.30$ Pa (visualization with ParaView software  \cite{ahrens2005paraview}).}
   \label{fig:olla112}
\end{figure}

Figure \ref{fig:olla112} represents the power density in cells of the FEM calculation where its value is found in the range of the colorbar ($0.1 \leq p_\Omega \leq 1000$ W/cm$^3$). The higher power density represented by darker regions concentrates around the FSLW slits, with larger values near the extremes (note the logarythmic scale). 
 The intense induction field in the slits, especially in the inside, is expected from theoretical considerations \cite{akhmetov2021on-the-influenc}.

  We obtain the following respective values of dissipated power in the plasma, back disk, FSLW cylinder and stainless steel ring due to electro-magnetic induction alone:
\begin{equation}
P_\mathrm{pl} = 9.27 \mbox{ kW; }P_\mathrm{bd} = 0.14 \mbox{ kW; }P_\mathrm{F} = 14.54 \mbox{ kW; }P_\mathrm{ss}= 2.35 \mbox{ kW.}
\label{ec:valores disipacion}
\end{equation}
We observe that the FSLW obtains most of the power absorbed by the metallic parts.

At present, the only possibility to confront these calculations with experimental data comes from calorimetry measurements. In the companion paper \cite{denizeau2024heating} we show a survey of SPIDER discharges with calorimetry data and first estimates of heat dissipation in several parts of the SPIDER sources during the past campaigns. In this survey it is found that discharges similar to the one used to produce figure \ref{fig:olla112} and the values Eq.~\ref{ec:valores disipacion}, dissipate around 15 kW through the cooling water circuits. We take this good agreement among the present model calculations and calorimetry estimates as a first check of the electro-magnetic models.

The measurements in \cite{denizeau2024heating} permit also an independent estimate of the PTE. Under a variety of common operational conditions (neutral gas pressure, generator power, magnetic filter field), the PTE from calorimetry data is roughly in the range 30--70\%, in acceptable agreement with the model electro-magnetic estimates presented in figure \ref{fig:eficiencias} for a reduced set of operating conditions. The measurements of cooling water temperature along with the power deposition data taken from the present electro-magnetic calculations permit also an estimate ot the FSLW temperature during operation. As shown in \cite{denizeau2024heating}, our choice of $\approx 50$ $^\circ$C to decide the copper conductivity is very reasonable for the studied operation conditions in SPIDER. It is also found that the decrease in copper conductivity in the hottest regions of the FSLW cylinder is small, below 4\%, which justifies our assumption of constant copper electrical conductivity. This aspect remains of interest, however, because it might not be true at generator powers well above 50 kW. Further work using the smaller facility MINION \cite{mario2024optimizing-the-} will be devoted to refine all these comparisons.

\section{\label{sec:discusion}Conclusions}

Previous 2D electromagnetic calculations of the inductive coupling between the RF coil and the plasma in the drivers of the SPIDER device \cite{lopezbruna20212Delectromagnet} have been extended to 3D including the coupling to metallic parts of the driver, like the Faraday shield cylinder \cite{lopezbruna2023threedimensional,lopezbruna2022Three-dimensional}. We use these new calculation tools to evaluate the power dissipated in different parts of the driver considering experimental SPIDER plasmas as input. In particular, the FSLW absorbs a considerable amount of the generator RF power per driver ---often near half of it--- depending on the copper conductivity (temperature) and, sensibly, the plasma profiles. The PTE is consequently low, approximately in the range 30--45\% for the plasmas under study ($20 \lesssim P_\mathrm{RF} \lesssim 50$ kW and different magnetic filter fields). Similar results have been found in comparable drivers (see \cite{briefi2022diagnostics-of-} and references therein). 
Hypothesizing denser plasmas in the regions of high induced electric field, as it might be the case with a proper configuration of the static magnetic field in the driver region, the efficiency can increase notably. The theoretical assessment of this hypothesis requires, however, solving a 3D transport problem not considered here.

We have identified  SPIDER discharges with available calorimetry and operational parameters similar to those of the campaign devoted to the plasma characterization. Despite the acknowledged lack of accuracy of calorimetry estimates in the present experiments, there is agreement between the calculated and measured power dissipated in the FSLW. In qualitative terms, the dissipated power in the FSLW cylinder is on the order of half the generator power per driver and the plasma transfer efficiency is often below 50\% in the present design of SPIDER drivers. New experiments are planned to refine these comparisons.

\section*{Acknowledgement}

This work has been carried out within the framework of the EUROfusion Consortium, funded by the European Union via the Euratom Research and Training Programme (Grant Agreement No 101052200 --- EUROfusion). Views and opinions expressed are however those of the author(s) only and do not necessarily reflect those of the European Union or the European Commission. Neither the European Union nor the European Commission can be held responsible for them.

The dedication of the whole NBTF Team is sincerely acknowledged.

\bibliographystyle{iopart-num} % iopart-num, amsalpha, IEEEtran
%\bibliography{/Users/dlb/Documents/NBTF}
\bibliography{/Users/Daniel/Documents/NBTF}

\providecommand{\newblock}{}
\begin{thebibliography}{10}
\expandafter\ifx\csname url\endcsname\relax
  \def\url#1{{\tt #1}}\fi
\expandafter\ifx\csname urlprefix\endcsname\relax\def\urlprefix{URL }\fi
\providecommand{\eprint}[2][]{\url{#2}}

\bibitem{Toigo2017The-PRIMA-Test-}
Toigo V, Piovan R, Bello S~D, Gaio E, Luchetta A, Pasqualotto R, Zaccaria P,
  Bigi M, Chitarin G, Marcuzzi D, Pomaro N, Serianni G, Agostinetti P, Agostini
  M, Antoni V, Aprile D, Baltador C, Barbisan M, Battistella M, Boldrin M,
  Brombin M, Palma M~D, Lorenzi A~D, Delogu R, Muri M~D, Fellin F, Ferro A,
  Fiorentin A, Gambetta G, Gnesotto F, Grando L, Jain P, Maistrello A, Manduchi
  G, Marconato N, Moresco M, Ocello E, Pavei M, Peruzzo S, Pilan N, Pimazzoni
  A, Recchia M, Rizzolo A, Rostagni G, Sartori E, Siragusa M, Sonato P,
  Sottocornola A, Spada E, Spagnolo S, Spolaore M, Taliercio C, Valente M,
  Veltri P, Zamengo A, Zaniol B, Zanotto L, Zaupa M, Boilson D, Graceffa J,
  Svensson L, Schunke B, Decamps H, Urbani M, Kushwah M, Chareyre J, Singh M,
  Bonicelli T, Agarici G, Garbuglia A, Masiello A, Paolucci F, Simon M,
  Bailly-Maitre L, Bragulat E, Gomez G, Gutierrez D, Mico G, Moreno J~F, Pilard
  V, Kashiwagi M, Hanada M, Tobari H, Watanabe K, Maejima T, Kojima A, Umeda N,
  Yamanaka H, Chakraborty A, Baruah U, Rotti C, Patel H, Nagaraju M~V, Singh
  N~P, Patel A, Dhola H, Raval B, Fantz U, Heinemann B, Kraus W, Hanke S, Hauer
  V, Ochoa S, Blatchford P, Chuilon B, Xue Y, Esch H~P~L~D, Hemsworth R, Croci
  G, Gorini G, Rebai M, Muraro A, Tardocchi M, Cavenago M, D'Arienzo M, Sandri
  S and Tonti A 2017 {\em New Journal of Physics\/} {\bf 19} 085004
  \urlprefix\url{https://iopscience.iop.org/article/10.1088/1367-2630/aa78e8}

\bibitem{iter-physics-expert-group-on-EP-1999chapter-6}
{ITER Physics Expert Group on Energetic Particles, Heating and Current Drive
  and ITER Physics Basis Editors} 1999 {\em Nuclear Fusion\/} {\bf 39} 2495
  \urlprefix\url{https://dx.doi.org/10.1088/0029-5515/39/12/306}

\bibitem{iter-physics-expert-group-on-EP2000chapter6}
{ITER Physics Expert Group on Energetic Particles, Heating and Current Drive
  and ITER Physics Basis Editors} 2000 {\em Nuclear Fusion\/} {\bf 40} 1429
  \urlprefix\url{https://dx.doi.org/10.1088/0029-5515/40/7/512}

\bibitem{mario2024optimizing-the-}
Mario I, Pimazzoni A, Sartori E, Pouradier-Duteil B, Sheperd A, Denizeau S,
  Casagrande R, Agnello R, Agostini M, Aprile D, Barbato P, Baseggio L,
  Battistella M, Berton G, Boldrin M, Brombin M, Candeloro V, Carraro M,
  Cinetto P, Bello S~D, Delogu R, Fadone M, Fellin F, Fincato M, Franchin L,
  Friso D, Grando L, Rosa A~L, Laterza B, L{\'o}pez-Bruna D, Magagna M, Maniero
  M, Marconato N, Pasqualotto R, Passalacqua G, Pavei M, Poggi C, Ravarotto D,
  Rigoni-Garola A, Romanato L, Rossetto F, Segalini B, Sonato P, Taliercio C,
  Toigo V, Tollin M, Ugoletti M, Vignando M, Zag{\'o}rski R, Zaniol B, Zaupa M,
  Zella D, Zerbetto E, Zucchetti S, Zuin E and Serianni G 2024 {\em Journal of
  Physics: Conference Series\/} {\bf 2743} 012041
  \urlprefix\url{https://dx.doi.org/10.1088/1742-6596/2743/1/012041}

\bibitem{toigo2021on-the-road-to-}
Toigo V, Marcuzzi D, Serianni G, Boldrin M, Chitarin G, Bello S~D, Grando L,
  Luchetta A, Pasqualotto R, Zaccaria P, Zanotto L, Agnello R, Agostinetti P,
  Agostini M, Antoni V, Aprile D, Barbisan M, Battistella M, Berton G, Bigi M,
  Brombin M, Candeloro V, Canton A, Casagrande R, Cavallini C, Cavazzana R,
  Cordaro L, Cruz N, Palma M~D, Dan M, {De Lorenzi} A, Delogu R, {De Muri} M,
  Denizeau S, Fadone M, Fellin F, Ferro A, Gaio E, Gasparini F, Gasparrini C,
  Gnesotto F, Jain P, Krastev P, Lopez-Bruna D, Lorenzini R, Maistrello A,
  Manduchi G, Manfrin S, Marconato N, Martines E, Martini G, Martini S, Milazzo
  R, Patton T, Pavei M, Peruzzo S, Pilan N, Pimazzoni A, Poggi C, Pomaro N,
  Pouradier-Duteil B, Recchia M, Rigoni-Garola A, Rizzolo A, Sartori E,
  Shepherd A, Siragusa M, Sonato P, Sottocornola A, Spada E, Spagnolo S,
  Spolaore M, Taliercio C, Terranova D, Tinti P, Tomsi{\v c} P, Trevisan L,
  Ugoletti M, Valente M, Vignando M, Zagorski R, Zamengo A, Zaniol B, Zaupa M,
  Zuin M, Cavenago M, Boilson D, Rotti C, Veltri P, Decamps H, Dremel M,
  Graceffa J, Geli F, Urbani M, Zacks J, Bonicelli T, Paolucci F, Garbuglia A,
  Agarici G, Gomez G, Gutierrez D, Kouzmenko G, Labate C, Masiello A, Mico G,
  Moreno J~F, Pilard V, Rousseau A, Simon M, Kashiwagi M, Tobari H, Watanabe K,
  Maejima T, Kojima A, Oshita E, Yamashita Y, Konno S, Singh M, Chakraborty A,
  Patel H, Singh N, Fantz U, Bonomo F, Cristofaro S, Heinemann B, Kraus W,
  Wimmer C, W{\"u}nderlich D, Fubiani G, Tsumori K, Croci G, Gorini G,
  McCormack O, Muraro A, Rebai M, Tardocchi M, Giacomelli L, Rigamonti D,
  Taccogna F, Bruno D, Rutigliano M, D'Arienzo M, Tonti A and Panin F 2021 {\em
  Fusion Engineering and Design\/} {\bf 168} 112622 ISSN 0920-3796
  \urlprefix\url{https://www.sciencedirect.com/science/article/pii/S0920379621003987}

\bibitem{jain2022investigation-o}
Jain P, Recchia M, Maistrello A and Gaio E 2022 {\em Plasma Physics and
  Controlled Fusion\/} {\bf 64} 095018
  \urlprefix\url{https://doi.org/10.1088/1361-6587/ac8617}

\bibitem{marconato2021an-optimized-an}
Marconato N, Brombin M, Pavei M, Tollin M, Baseggio L, Fincato M, Franchin L,
  Maistrello A and Serianni G 2021 {\em Fusion Engineering and Design\/} {\bf
  166} 112281 ISSN 0920-3796
  \urlprefix\url{https://www.sciencedirect.com/science/article/pii/S0920379621000570}

\bibitem{sartori2023highlights-of-r}
Sartori E, Agnello R, Agostini M, Barbisan M, Bigi M, Boldrin M, Brombin M,
  Candeloro V, Casagrande R, Bello S~D, Dan M, Duteil B~P, Fadone M, Grando L,
  Jain P, Maistrello A, Mario I, Pasqualotto R, Pavei M, Pimazzoni A, Poggi C,
  Rizzolo A, Shepherd A, Ugoletti M, Veltri P, Zaniol B, Agostinetti P, Aprile
  D, Berton G, Cavallini C, Cavenago M, Chitarin G, Croci G, Delogu R, Muri
  M~D, Nardi M~D, Denizeau S, Fellin F, Ferro A, Gaio E, Gasparrini C, Luchetta
  A, Lunardon F, Manduchi G, Marconato N, Marcuzzi D, McCormack O, Milazzo R,
  Muraro A, Patton T, Pilan N, Recchia M, Rigoni-Garola A, Santoro F, Segalini
  B, Siragusa M, Spolaore M, Taliercio C, Toigo V, Zaccaria P, Zagorski R,
  Zanotto L, Zaupa M, Zuin M and Serianni G 2023 {\em Journal of
  Instrumentation\/} {\bf 18} C09001
  \urlprefix\url{https://dx.doi.org/10.1088/1748-0221/18/09/C09001}

\bibitem{marconato2023integration-of-}
Marconato N, Berton G, Candeloro V, Sartori E, Segalini B and Serianni G 2023
  {\em Fusion Engineering and Design\/} {\bf 193} 113805 ISSN 0920-3796
  \urlprefix\url{https://www.sciencedirect.com/science/article/pii/S0920379623003873}

\bibitem{briefi2022diagnostics-of-}
Briefi S, Zielke D, Rauner D and Fantz U 2022 {\em Review of Scientific
  Instruments\/} {\bf 93} 023501 ISSN 0034-6748 (\textit{Preprint}
  \eprint{https://pubs.aip.org/aip/rsi/article-pdf/doi/10.1063/5.0077934/16714888/023501\_1\_online.pdf})
  \urlprefix\url{https://doi.org/10.1063/5.0077934}

\bibitem{lopezbruna2023threedimensional}
L\'opez-Bruna D, Recchia M, Jain P, Predebon I, Denizeau S and Rosa A~L 2023
  Three-dimensional calculations of the inductive coupling between
  radio-frequency waves and plasma in the drivers of the {SPIDER} device
  arXiv:2310.09878 [physics.plasm-ph]
  \urlprefix\url{https://doi.org/10.48550/arXiv.2310.09878}

\bibitem{zagorski20222-d-fluid-model}
Zag\'orski R, Sartori E and Serianni G 2022 {\em IEEE Transactions on Plasma
  Science\/}  1--7 ISSN 1939-9375

\bibitem{zagorski20232d-simulations}
Zag{\'o}rski R, L{\'o}pez-Bruna D, Sartori E and Serianni G 2023 {\em Fusion
  Engineering and Design\/} {\bf 188} 113427 ISSN 0920-3796
  \urlprefix\url{https://www.sciencedirect.com/science/article/pii/S092037962300011X}

\bibitem{lopezbruna2022Three-dimensional}
L\'opez-Bruna D, Recchia M, Jain P and Predebon I 2022 Three-dimensional
  calculations of the inductive coupling between radio-frequency waves and
  plasma in the drivers of the {SPIDER} device {\em 8th International symposium
  on Negative Ions, Beams and Sources - NIBS'22\/} (Padova, Italy)

\bibitem{chen2018electromagnetic}
Chen P, Li D, Chen D, Song F, Zuo C, Zhao P and Lei G 2018 {\em AIP Conference
  Proceedings\/} {\bf 2052} 040018 ISSN 0094-243X (\textit{Preprint}
  \eprint{https://pubs.aip.org/aip/acp/article-pdf/doi/10.1063/1.5083752/13167933/040018\_1\_online.pdf})
  \urlprefix\url{https://doi.org/10.1063/1.5083752}

\bibitem{wang2022design-optimiza}
Wang N, Liu Z, Xie Y, Wang J, Wei J, Gu Y, Cui Q, Xie Y and Hu C 2022 {\em
  Fusion Engineering and Design\/} {\bf 183} 113272 ISSN 0920-3796
  \urlprefix\url{https://www.sciencedirect.com/science/article/pii/S0920379622002642}

\bibitem{denizeau2024heating}
Denizeau S, L\'opez-Bruna D, Agostinetti P, Berton G, Rosa A~L and Pavei M 2024
  {\em Plasma Sources Science and Technology (submitted)\/}

\bibitem{Jain2018Evaluation-of-p}
Jain P, Recchia M, Cavenago M, Fantz U, Gaio E, Kraus W, Maistrello A and
  Veltri P 2018 {\em Plasma Physics and Controlled Fusion\/} {\bf 60} 045007
  \urlprefix\url{https://doi.org/10.1088/1361-6587/aaab19}

\bibitem{Jain2018Improved-Method}
{Jain} P, {Recchia} M, {Veltri} P, {Cavenago} M, {Maistrello} A and {Gaio} E
  2018 {\em IEEE Access\/} {\bf 6} 29665--29676 ISSN 2169-3536

\bibitem{fenicsproject}
{FEniCS} project \urlprefix\url{https://fenicsproject.org}

\bibitem{Logg2012Automated-Solut}
Logg A, Mardal K~A, Wells G~N {\em et~al.\/} 2012 {\em Automated Solution of
  Differential Equations by the Finite Element Method\/} (Springer) ISBN
  978-3-642-23098-1

\bibitem{Langtangen2017Solving-PDEs-in}
Langtangen H~P and Logg A 2017 {\em Solving PDEs in Python\/} (Springer) ISBN
  978-3-319-52461-0

\bibitem{Alnaes2012UFL}
Aln\ae{}s M~S 2012 {\em UFL: a Finite Element Form Language\/} (Springer)
  chap~17

\bibitem{Logg2010DOLFIN}
Logg A and Wells G~N 2010 {\em ACM Transactions on Mathematical Software\/}
  {\bf 37}

\bibitem{Logg2012DOLFIN}
Logg A, Wells G~N and Hake J 2012 {\em DOLFIN: a C++/Python Finite Element
  Library\/} (Springer) chap~10

\bibitem{Geuzaine2009Gmsh:-A-3-D-fin}
Geuzaine C and Remacle J~F 2009 {\em International Journal for Numerical
  Methods in Engineering\/} {\bf 79} 1309--1331 (\textit{Preprint}
  \eprint{https://onlinelibrary.wiley.com/doi/pdf/10.1002/nme.2579})
  \urlprefix\url{https://onlinelibrary.wiley.com/doi/abs/10.1002/nme.2579}

\bibitem{lopezbruna20212Delectromagnet}
L\'opez-Bruna D, Jain P, Recchia M, Zaniol B, Sartori E, Poggi C, Candeloro V,
  Serianni G and Veltri P 2023 {2D} electromagnetic simulations of {RF} heating
  via inductive coupling in the {SPIDER} device arXiv:2305.09395v1
  [physics.plasm-ph] \urlprefix\url{https://doi.org/10.48550/arXiv.2305.09395}

\bibitem{sartori2021development-of-}
Sartori E, Brombin M, Laterza B, Zuin M, Cavazzana R, Cervaro V, {Degli
  Agostini} F, Fadone M, Fasolo D, Grando L, Jain P, Kisaki M, Maistrello A,
  Moro G, Pimazzoni A, Poggi C, Segalini B, Shepherd A, Spolaore M, Taliercio
  C, Tollin M, Ugoletti M, Veltri P, Zamengo A and Serianni G 2021 {\em Fusion
  Engineering and Design\/} {\bf 169} 112424 ISSN 0920-3796
  \urlprefix\url{https://www.sciencedirect.com/science/article/pii/S0920379621002003}

\bibitem{Vahedi1995Analytic-model-}
Vahedi V, Lieberman M~A, DiPeso G, Rognlien T~D and Hewett D 1995 {\em Journal
  of Applied Physics\/} {\bf 78} 1446--1458 ISSN 1089-7550
  \urlprefix\url{http://dx.doi.org/10.1063/1.360723}

\bibitem{tuszewski1997inductive-elect}
Tuszewski M 1997 {\em Physics of Plasmas\/} {\bf 4} 1922--1928
  (\textit{Preprint} \eprint{https://doi.org/10.1063/1.872335})
  \urlprefix\url{https://doi.org/10.1063/1.872335}

\bibitem{agostinetti2023innovative}
Agostinetti P, Benedetti E, Bonifetto R, Bonesso M, Calabr{\`o} G, Cavenago M,
  Crisanti F, Bello S~D, Palma M~D, D'Ambrosio D, Dima R, Favero G, Ferro A,
  Fincato M, Grando L, Granucci G, Lombroni R, Marsilio R, andT Patton A~M,
  Pepato A, Raffaelli F, Rebesan P, Recchia M, Ripani M, Romano A, Sartori E,
  Scarpari M, Variale V, Ventura G, Veronese F, Zanino R, Zappatore A and
  Zavaris G 2023 Innovative concepts in the dtt neutral beam injector {\em
  Procs. of the IEEE Symposium on Fusion Engineering (SOFE), Oxford\/}

\bibitem{ahrens2005paraview}
Ahrens J, Geveci B and Law C 2005 Paraview: An end-user tool for large-data
  visualization {\em Visualization Handbook\/} ed Hansen C~D and Johnson C~R
  (Burlington: Butterworth-Heinemann) pp 717--731 ISBN 978-0-12-387582-2
  \urlprefix\url{https://www.sciencedirect.com/science/article/pii/B9780123875822500381}

\bibitem{akhmetov2021on-the-influenc}
Akhmetov T~D, Ivanov A~A and Shikhovtsev I~V 2021 {\em Plasma Physics
  Reports\/} {\bf 47} 732--741
  \urlprefix\url{https://doi.org/10.1134/S1063780X21070011}

\end{thebibliography}

\end{document}